\documentclass[a4paper,11pt]{article}
\pdfoutput=1 % if your are submitting a pdflatex (i.e. if you have
\usepackage{jinstpub} % for details on the use of the package, please
\usepackage{lineno}
\usepackage{wasysym}
\usepackage{hyperref}
\usepackage{subcaption}
\usepackage{wrapfig}

\title{Optical tomography diagnostic for study of neutral plasma component in the gas dynamic trap}

\author[1]{A. Lizunov\note{Corresponding author.}, }
\author{A. Khilchenko, }
\author{D. Moiseev} 
\author{ and P. Zubarev}

\affiliation{Budker Institute of nuclear physics,\\630090 Novosibirsk, Russia}

\emailAdd{lizunov@inp.nsk.su}

\abstract{The optical diagnostic observing D-$\alpha$ line emission along multiple chords in the boundary region close to the plasma absorber, is recently installed at the gas dynamic trap. The implemented pattern of viewing lines is suitable for a tomographic reconstruction of local emissivity profiles, although steps towards increasing the channel count and extending of angular plasma coverage must be taken. The optical registration system of a modular design uses avalanche photodiodes with wideband amplifiers for a large signal dynamic range and the effective time resolution of $1 \mu s$. The iterative backward projection technique based on the maximum likelihood principle, demonstrates an acceptable computation accuracy. Images of plasma evolution in the cross section were obtained. Tools for the correlation analysis were also developed and first results of study of the plasma turbulence are presented. }


\arxivnumber{1234.56789} % only if you have one

\begin{document}
\maketitle
\flushbottom

\section{Introduction}
\label{sec:intro}
Linear open-ended systems have direct contact of confined plasmas with the end wall due to the magnetic field geometry. The axial heat conductivity is therefore a major issue which must be addressed in an improved confinement concept basing on the linear "magnetic bottle" field configuration. As one of them, the axially symmetric gas dynamic trap (GDT) \cite{gdt-review-ppcf} utilizes several techniques to increase the lifetime of particle and energy in the central cell. Strongly expanding magnetic flux beyond the mirror prevents a flyover penetration of cold electrons emitted by the wall back to central plasma thus allowing for a radical depression of heat conductivity comparing to the classical Spitzer \cite{spitzer} case. Study of axial transport \cite{nf-axconf-2020} is one of ongoing physical objectives of the GDT research program. In high-beta regimes \cite{highbeta_prl} with electron cyclotron resonance heating (ECRH) \cite{ecrh_prl}, extra efforts must be spent to stabilize the plasma against magneto-hydrodynamic (MHD) perturbations. This line of experimental and theoretical activities has led to establishing of the so called vortex confinement \cite{vortexconf} method of MHD modes suppression which is now a standard attribute of experimental scenarios in GDT. A specific plasma equilibrium called vortex, is driven by biasing radial limiters and plasma absorber sections. It is also realized that the approach is viable if the particle density is above the certain threshold in the expander region. Normally this density is maintained by the hydrogen or deuterium puff in the expander tank. The neutral plasma component formed both by the gas puff and plasma flux neutralisation on the end wall, therefore represents a significant fraction bound to the physics of ion and electron transport in the expander. A productive way of study is detection of light emitted along multiple lines of sight (LOS) distributed across the plasma diameter. Then a tomography reconstruction can be used to obtain the local emissivity distribution. Various tomography diagnostics in optical to soft X-ray range are being widely used in experiments with magnetically confined plasmas for decades \cite{tomo-1, tomo-2, tomo-3, tomo-4, tomo-5, tomo-6, tomo-7, tomo-8, tomo-9}. The paper discusses the recently developed visible range optical diagnostic, which is installed in the GDT expander to monitor dynamics of atomic plasma component spatial profile.

\section{Diagnostic description}
\label{diagnostic}
\begin{figure}[htbp]
\centering 
\includegraphics[width=.9\textwidth]{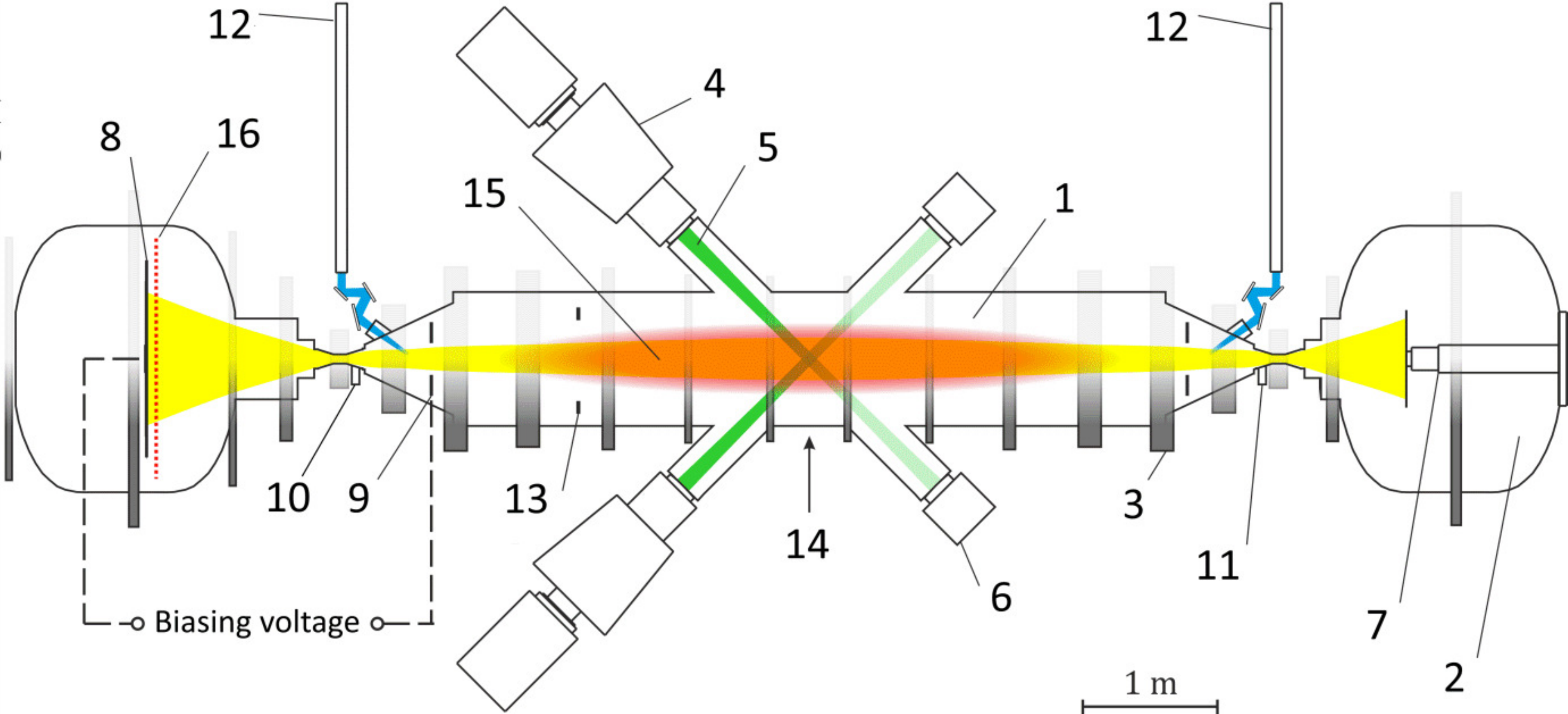}
\caption{\label{fig:gdt} Gas dynamic trap schematic: 1 -- central cell, 2 -- right expander tank, 3 -- magnetic coil of central solenoid, 4 -- atomic beam injector, 5 -- deuterium beam, 6 -- beam dump, 7 -- arc discharge plasma source, 8 -- plasma dump in the left expander tank, 9 -- radial limiter, 10 -- left gas box, 11 -- right gas box,  12 -- waveguides of ECRH system, 13 -- diamagnetic loop, 14 -- Thomson scattering diagnostic, 15 -- hot-ion plasma, 16 -- plane of observation of tomography diagnostic.}
\end{figure}

\begin{figure}[htbp]
\centering 
\includegraphics[width=.7\textwidth]{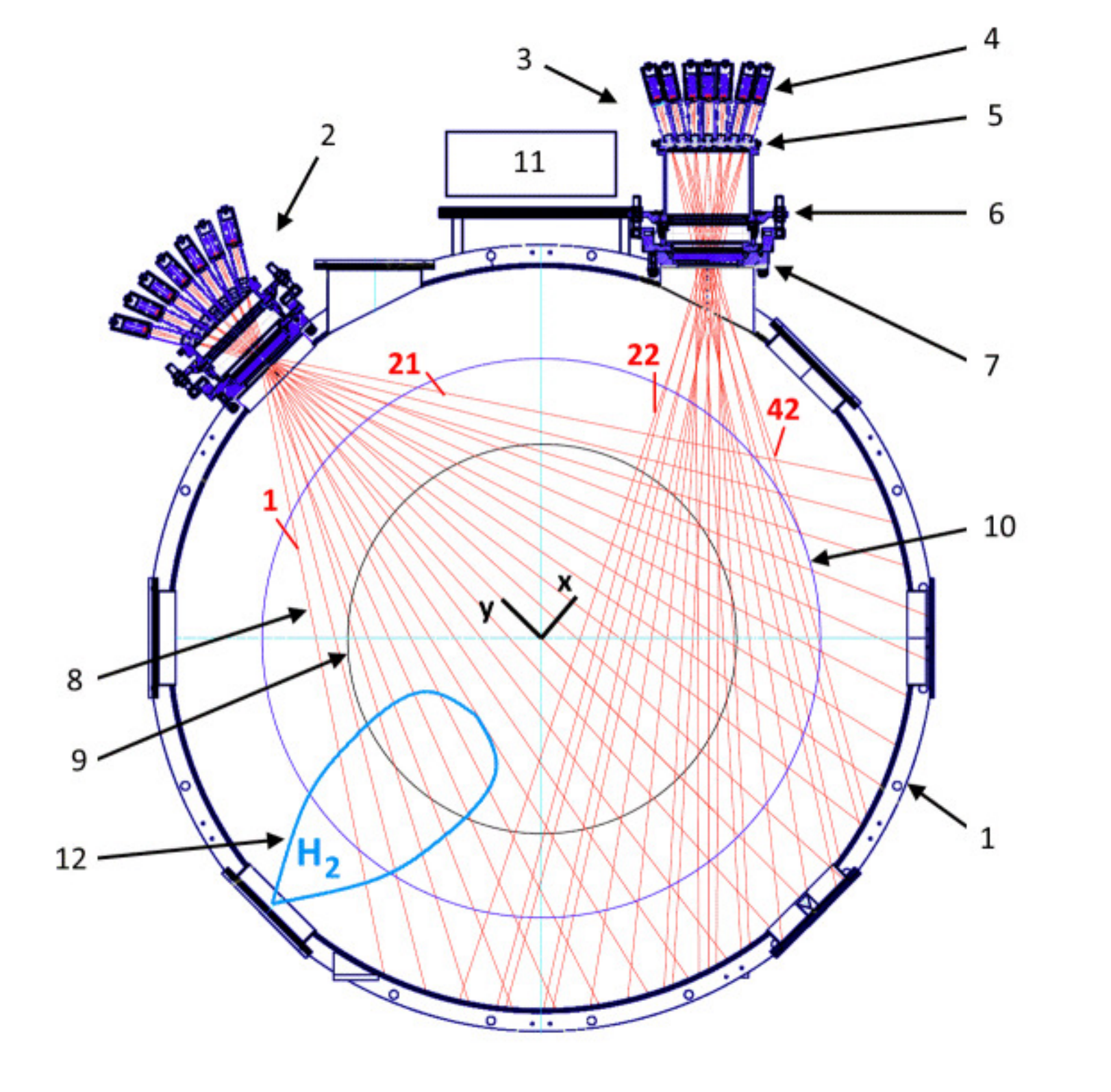}
\caption{\label{fig:w-exp_WIDA} Diagnostic layout: 1 -- left GDT expander tank, 2 -- diagnostic bundle-1, 3 -- diagnostic bundle-2, 4 -- detector module, 5 -- interference filter nest, 6 -- angle alignment unit, 7 -- vacuum window, 8 -- line of sight, 9 -- projection of radial limiter, 10 -- maximum observable plasma size, 11 -- electronics box, 12 -- gas puff cloud.}
\end{figure}

Figure~\ref{fig:gdt} shows the gas dynamic trap layout.
The diagnostic LOS are arranged in the plane {\bf 16}, which is offset at 50~mm from the plasma absorber surface {\bf 6}. The viewing geometry is presented in Figure~\ref{fig:w-exp_WIDA}.
In its current state, the diagnostic has two LOS bundles of a fan beam structure both. The developer team bears plans to install several (four) bundles similar to the bundle-1 to cover all angles nearly uniformly. Unfortunately the vacuum port availability was a severe limiting factor during this period of time, which has defined a suboptimal plasma coverage with LOS. The bundle-2 is installed in the tangent view port having accordingly smaller aperture angle. Undoubtedly this LOS pattern would sacrifice the Abel inversion accuracy which degree is estimated below in the paper. Up to now, the optical system counts 42 LOS numbered as shown in Figure~\ref{fig:w-exp_WIDA}. The whole optical registration system is comprised of 14 identical detector modules with three avalanche photodiodes (APD) sharing the same lens and interference filter. The optical and mechanical module design is discussed in the previous paper describing the prototype diagnostic \cite{WIDA_first} along with consideration of the two-stage transimpedance operational amplifier (OPA) and the procedure of absolute intensity calibration. The current version uses narrowband interference filters centred at 656.2~nm with the FWHM of 1~nm to observe both H-$\alpha$ and D-$\alpha$ emitted by the plasma majority and gas puffed into the expander. In principle, one can switch to an impurity line (within the wavelength range of a respectable APD quantum efficiency) simply replacing the set of filters. Filter-lens interchangeable assemblies are mounted in the aluminium body part 5 (see Figure~\ref{fig:w-exp_WIDA}) which is provided with water cooling flowing in the recirculating chiller system for the working temperature set point of $20^\circ$C. Detector modules are attached to this cooled part having approximately the same temperature $T_{APD} = 20\pm2^\circ$C. Prior to installation, each detector module is tuned to the optimal APD bias voltage individually and calibrated at this voltage. This aim of this procedure is to adjust the avalanche gain in order to reach the maximum signal-to-noise ratio (SNR). On incident light signals close to what expected in GDT measurements, all acquisitions channels have shown $SNR = 85\div120$ within the bandwidth of 0...5~MHz. The APD-OPA performance parameters are competitive to that reported for similar devices developed for high-speed low light plasma diagnostics, see for example \cite{APD-OPA}. Lab tests have shown that both the SNR of the APD-OPA circuit and therefore the signal acquisition system effective dynamic range are defined by the Poisson shot noise and the APD excess noise but not the amplifier noise.  On the machine, two custom made precision multi-channel high voltage power supplies serves detector units in the bundle-1 and bundle-2. The output voltage is traced and recorded for every channel indicating that the set voltage stability (including both the ripple and the long-term drift) is $\cong10^{-4}$. Power supply units are placed in the electronics box {\bf 11} (see Figure~\ref{fig:w-exp_WIDA}), where signal loggers are homed as well. A compact arrangement of hardware minimises cable lengths and permits to have all electronics be grounded in the single point at the GDT port flange. For the described tomography diagnostic, we used the BINP made synchronous multi-channel ADC system with the sampling frequency of $50$~MSample/s and 12~bit vertical resolution. The recorder system has the System-on-Chip controller with the integrated Ethernet onboard and collected data is available on the remote server via the dedicated high-speed transfer protocol over TCP. In all measurements, the data acquisition system was delivering the actual time resolution of 1~$\mu s$ with the oversampling of 50~samples per data point. Such a provision is effective for smoothing out shot noise and APD excess noise at the price of bigger data rate through the connection line. In future ADC designs, an onboard Field Programmable Gate Array (FPGA) will be introduced for online data processing and reduction. Table~\ref{WIDA_param} summarises the main parameters of the tomography diagnostic on GDT.
\begin{table}[htbp]
\centering
\caption{\label{WIDA_param} Main parameters of the optical tomography diagnostic.}
\smallskip
\begin{tabular}{| l | l |}
\hline
\emph{Viewing geometry} & \\
\hline
Inteference filters & Centre 656.2 nm, FWHM 1 nm (1 filter for 3 LOS) \\
Optical lens & Single lens $\diameter$25.4 mm for 3 LOS \\
Current number of LOS & 42 \\
Maximum covered plasma radius & 20.3 cm (projection to central GDT plane) \\
\hline
\emph{Signal registration system} & \\
\hline
Sensor & APD Hamamatsu S12053-10 $\diameter$1 mm \\
APD quantum efficiency & 0.7 at 650 nm \\
Avalanche multiplication & $50 \div 70$ \\
Amplifier responsivity & $\cong 2\cdot10^5 V/A$ \\
Amplifier bandwidth & 0...5 MHz \\
ADC sampling rate &  50 MSample/s \\
ADC vertical resolution & 12 bit \\
SNR & $85\div120$ \\
Effective time resolution & $1\mu s$ \\
\hline
\end{tabular}
\end{table}

{\bf Measurement of absolute intensity.} With the absolute calibration made, each LOS delivers the optical power in Watts or number of collected photons per second per unity of volume. The ultimate physical task is to recover the local density of atomic hydrogen and deuterium fractions. This task would require a collisional-radiative model with a number of inputs. First of all, the data on electron temperature and density is necessary. The plasma absorber has a set of probes acquiring the total particle flux, ion current, incident energy and electron temperature. Upon the commissioning of these probes distributed across the entire absorber surface, arrays of parameters will be enabled for modelling of particle interactions in the sheath and expander volume. The theoretical model itself is under development as well. Not having a comprehensive description of the plasma equilibrium, one can derive a simple estimation of the excited state density
$n_3^* = \epsilon \tau_{rad}\cdot k_{3-2}$, where
$n_3^*$ is the density of hydrogen or deuterium atoms in the $n=3$ state, $\epsilon$ is the calculated local emissivity ($cm^{-3}s^{-1}$), $\tau_{rad}$ is the radiative decay time of the $n=3$ state and the $k_{3-2}$ is the branching ratio for H-$\alpha$ optical transition. 

\begin{figure}[htbp]
	\centering
	\begin{subfigure}{.45\textwidth}
		\includegraphics[width=\textwidth, trim=0 30 20 10,clip]{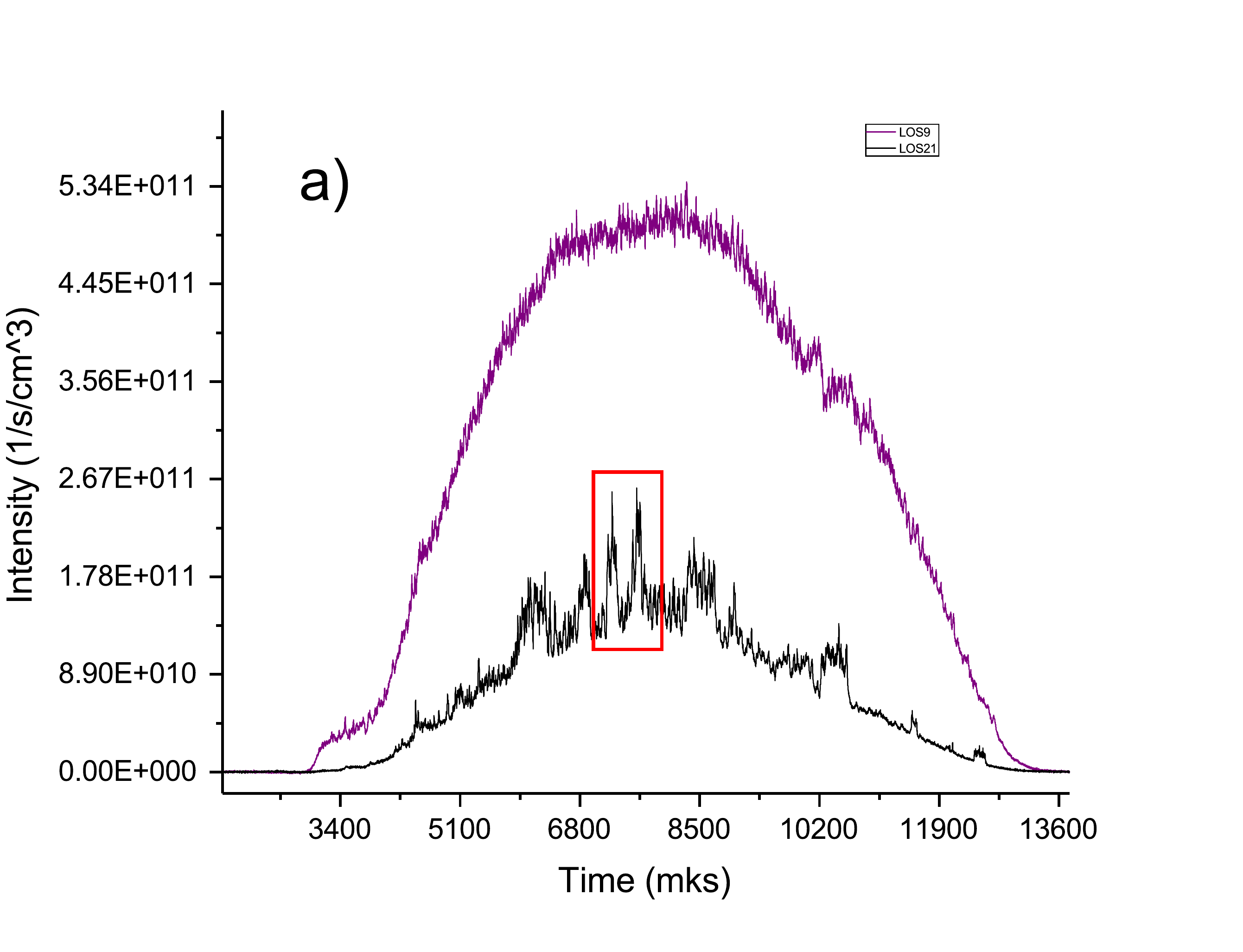}
		\caption{\label{LOS_signals} Signals from LOS-9 (bundle-1 central) and LOS-21 (bundle-1 edge) in GDT shot 49617 illustrating jumps and oscillations of intensity. }
	\end{subfigure}
	\qquad
	\begin{subfigure}{.45\textwidth}
		\includegraphics[width=\textwidth, trim=0 30 20 10,clip]{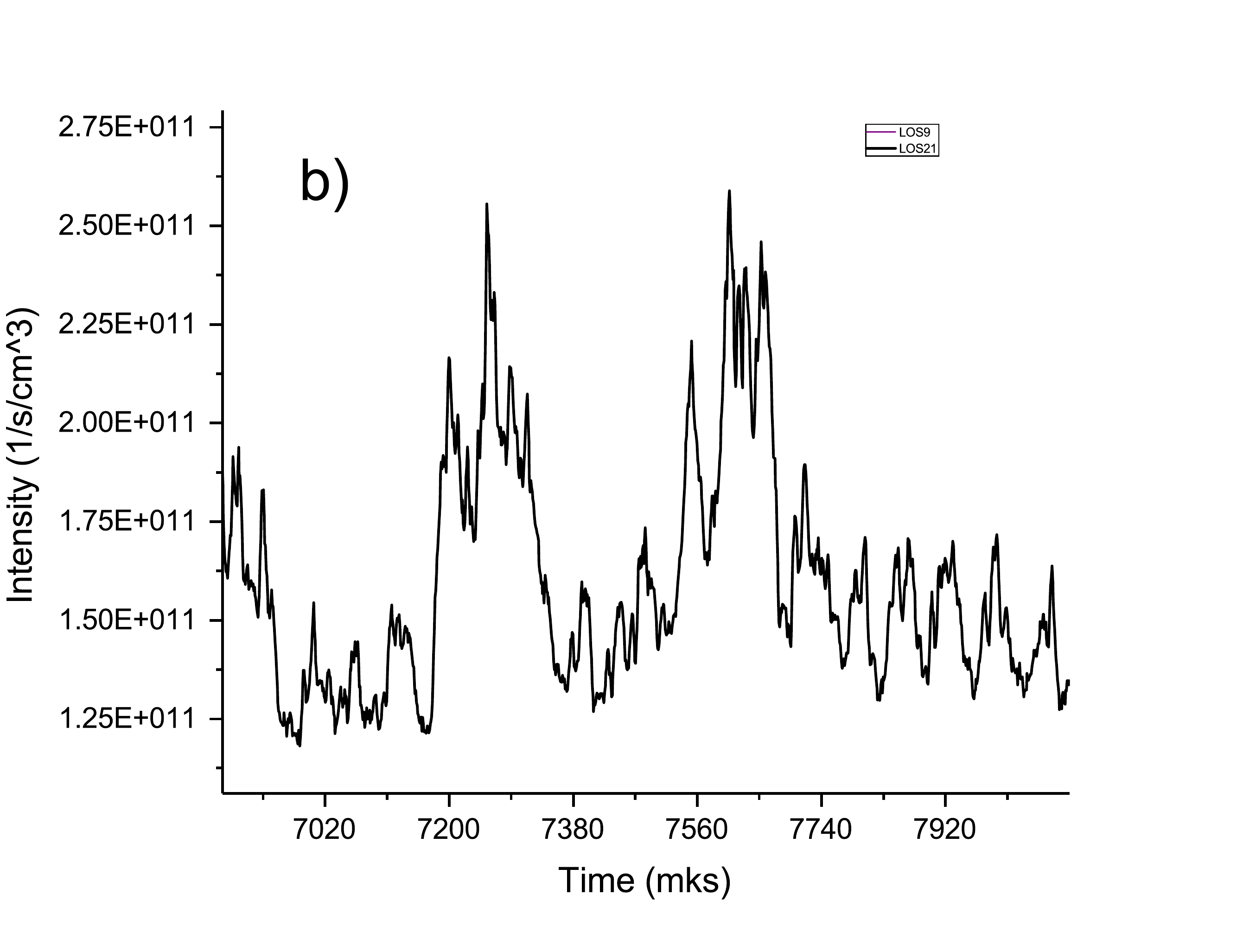}
		\caption{\label{LOS_signals_scaled}  Signal of LOS-21 scaled (boxed piece in Figure a). }
	\end{subfigure}
	\caption{Examples of signals of intensity integrated along LOS. }
\end{figure}

Figures~\ref{LOS_signals} and \ref{LOS_signals_scaled} show typical acquired signals of intensity integrated along the LOS, normalised on the observation solid angle and the light collection volume.
Most plasma confinement regimes are accompanied with intensity oscillations visible in Figures~\ref{LOS_signals} and \ref{LOS_signals_scaled}. Oscillations are more prominent in edge plasmas. First attempt of the time-domain and space-domain analysis of oscillations is discussed in the Section~\ref{Correlations}. Note that the detector noise is approximately two orders of magnitude smaller than small-scale fluctuations in Figure~\ref{LOS_signals_scaled}. 

\section{Method of tomographic reconstruction}
\label{Tomo_method}
Given pattern of LOS provides insufficient data and so the geometry matrix connecting the unknown emissivity distribution with the measured one, is underdetermined in our case.  Generally speaking, the mathematical problem can be qualified as ill-posed. Reconstruction techniques operating with the Fourier transform and matrix inversion methods \cite{calc_method_Fourier-1, calc_method_Fourier-2, calc_method_FBP-1, calc_method_FBP-2, calc_method_FBP-3, calc_method_matrix-1, calc_method_matrix-2}, face difficulties. For such underdetermined systems, regularisation or iterative fitting methods can be used \cite{calc_method_SVD-1, calc_method_SVD-2}. For our purpose, we have employed the iterative algorithm of maximum likelihood with mathematical expectation maximisation (ML-EM) \cite{calc_method_FBP-2, calc_method_FBP-3, calc_method_ML-EM-1, calc_method_ML-EM-2}. This algebraic method exhibits great advantages of good handling of sparse sampling, cone or fan beams and a limited viewing angle \cite{calc_method_ML-EM-1, calc_method_ML-EM-2}. No symmetry assumptions involved in the scheme, which is mandatory. However, this approach is a relatively demanding to computational resources.

The iterative sequence of the objective function maximisation is performed via the equation
\begin{equation}
\label{ML-EM}
\epsilon_i^{(n+1)} = \epsilon_i^{(n)} \cdot \frac{1}{\sum_{k}W_{ik}} \cdot \sum_{k}W_{ik}\frac{J_k}{\sum_{l}W_{lk}\epsilon_l^{(n)}},
\end{equation}
where $\epsilon_i^{(n)}$ is the emissivity in the grid cell-$i$ on the iteration step $n$, $J_k$ is the line integrated intensity of the LOS-$k$, $W_{ij}$ is the weight matrix defined by the geometry only. The computation starts from the flat distribution as an initial guess. On every step, (1) the forward projection is performed from the estimate as $J_i^{sim (n)} = \sum_k W_{ki} \epsilon_k^{(n)}$, then (2) it is compared to the measured one via the ratio $R_i$ = $J_i^{sim (n)} / J_i$ and then (3) the previous estimate is improved using \eqref{ML-EM}. The sequence iterates upon convergence. After some optimisation, the square XY grid of $20\times20$ cells was adapted, where $X, Y \in [-800, 800]$~mm. The axis direction is indicated in Figure~\ref{fig:w-exp_WIDA}. The computation box is projected to the GDT central plane radius of $r_0^{(max)} \approx 20.3$~cm exceeding the radial limiter radius of 14~cm. 

\begin{figure}[htbp]
	\centering
	\begin{subfigure}{.45\textwidth}
		\includegraphics[width=\textwidth, trim=10 0 10 0,clip]{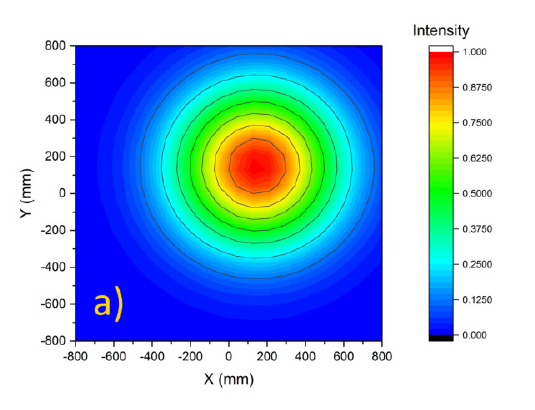}
		\caption{\label{I_model} Example of synthetic Gaussian profile used for validation, centre: $X_0 = Y_0 = 150$~mm. }
	\end{subfigure}
	\qquad
	\begin{subfigure}{.45\textwidth}
		\includegraphics[width=\textwidth, trim=10 0 10 0,clip]{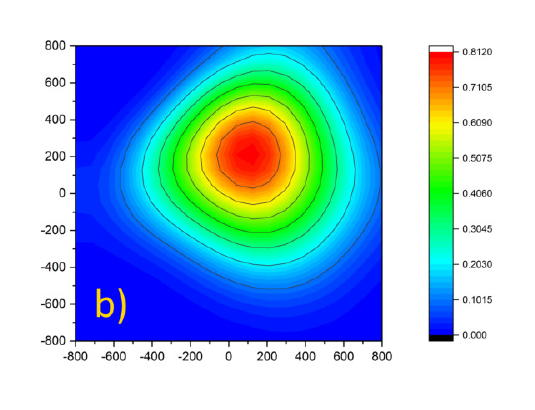}
		\caption{\label{I_model_recovered} Backward reconstruction of synthetic profile in Figure a). }
	\end{subfigure}
	\caption{Example of validation of tomography reconstruction algorithm on synthetic profiles.}
\end{figure}

The realised calculation scheme can be validated on a synthetic emissivity profile. It is used to simulate line-integrated intensities and then the latter is passed to the reconstruction code. Comparison between the synthetic profile and the calculated one is illustrated in Figures~\ref{I_model} and \ref{I_model_recovered}. 
Besides the de-centred Gaussian profile shown, other model shapes were also processed yielding similar results. A trapezoid distortion is present in the computed image as one would expect from the current limited angle optical registration system. We believe reconstruction artefacts will be significantly depressed in the planned advanced setup with four LOS beams and more even angular separation. Considering measurements taken in GDT plasmas so far, a certain caution must be adhered to quantitative conclusions regarding spatial distributions of the local emissivity. Nevertheless, reconstructed profiles are useful to observe their behaviour during the shot.

\section{Dynamics of emissivity profile in GDT expander plasmas} 
\label{I_profiles}
\begin{figure}[htbp]
\centering % \begin{center}/\end{center} takes some additional vertical space
\includegraphics[width=.9\textwidth]{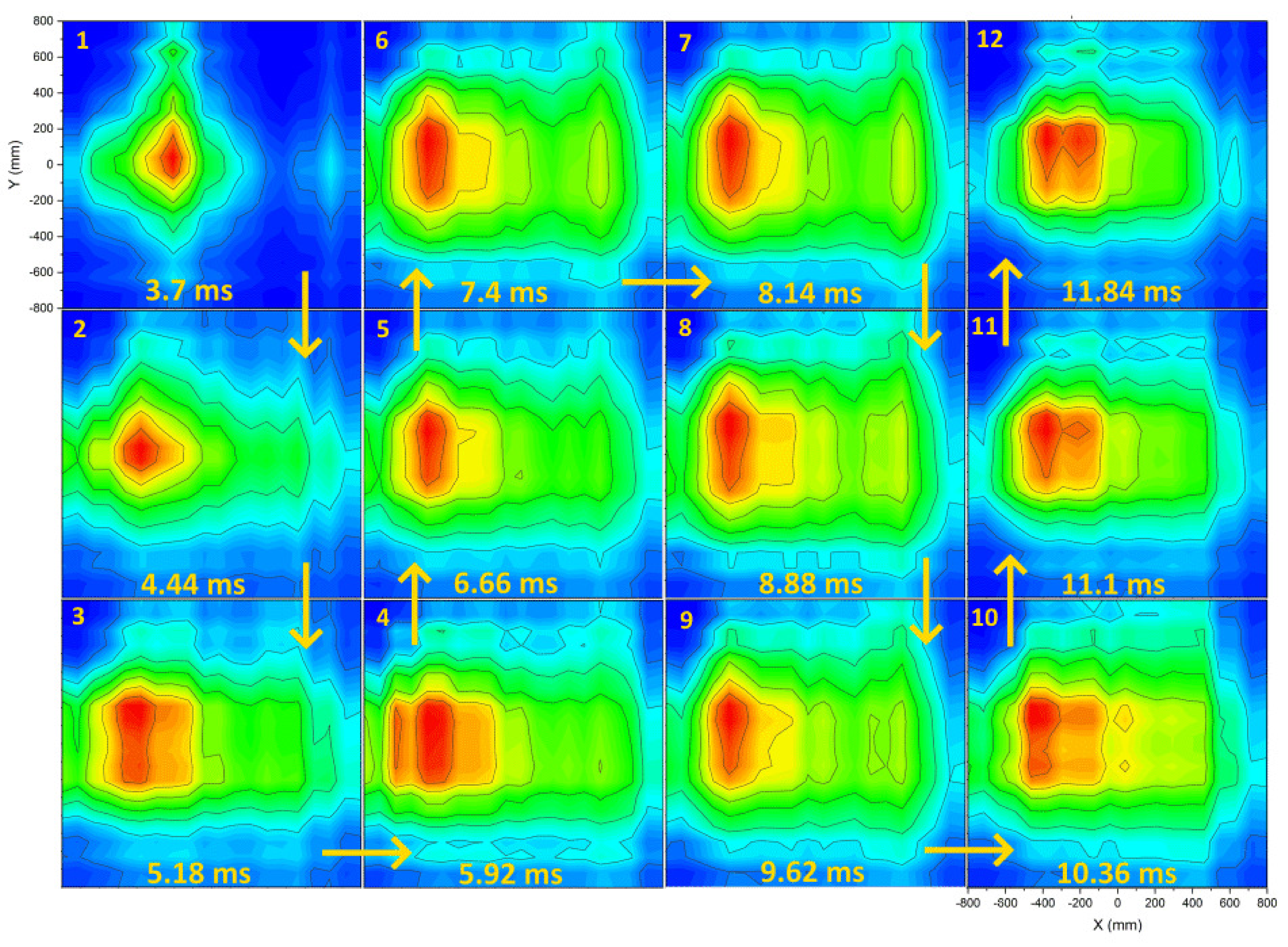}
\caption{\label{I_profiles_46617}  D-$\alpha$ emissivity profile dynamics in GDT shot 46617.}
%\qquad
\end{figure}

\begin{figure}[htbp]
\centering % \begin{center}/\end{center} takes some additional vertical space
\includegraphics[width=.45\textwidth,origin=c, trim=10 0 10 10, clip]{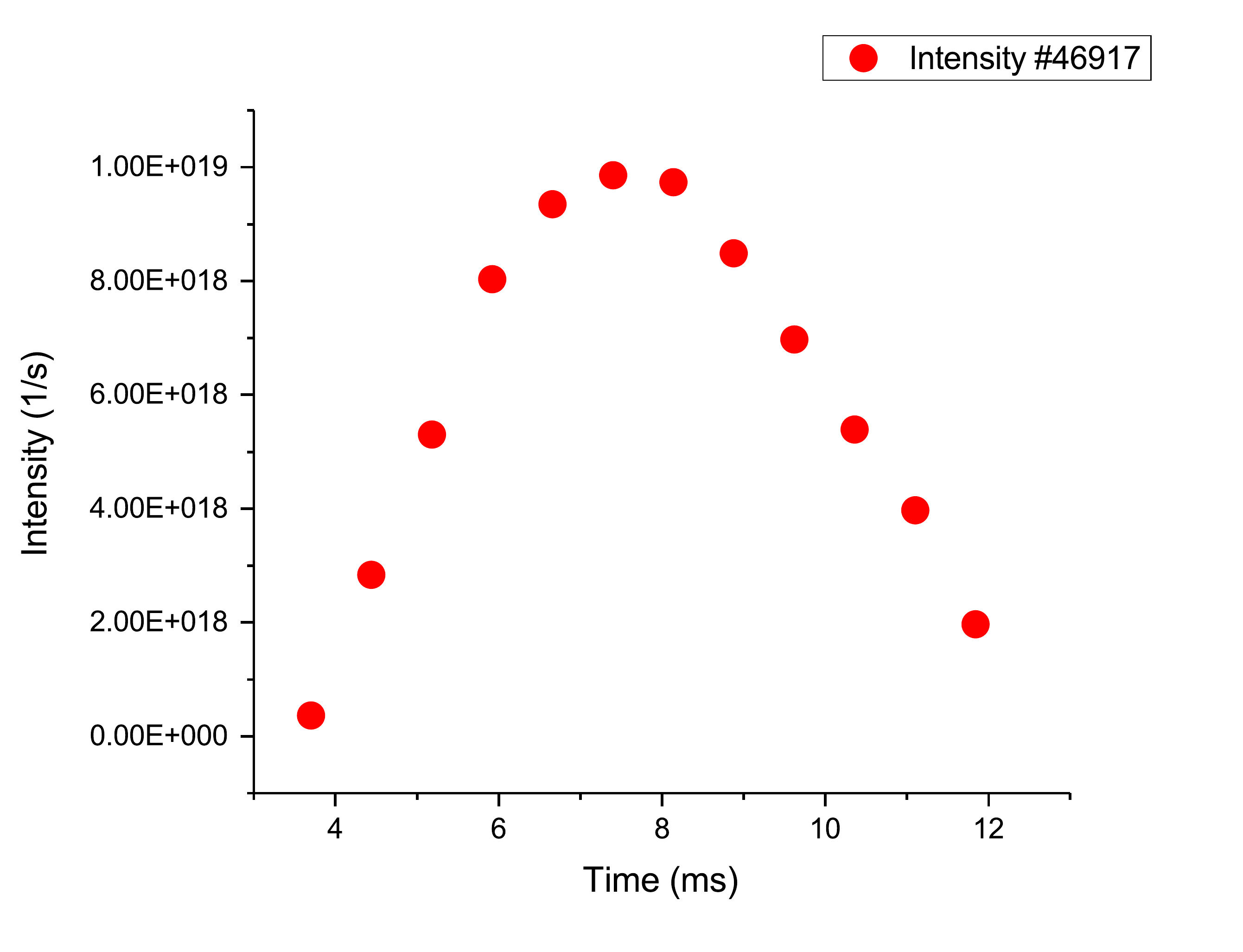}
\caption{\label{I_vs_t_46617} Dynamics of total D-$\alpha$ intensity in GDT shot 46617. }
\end{figure}

Figure~\ref{I_profiles_46617} demonstrates the series of reconstructed emissivity images in the GDT shot 46617 covering the plasma heating phase and decay.
Each frame's duration is $\tau_{exp} = 1 \mu s$ (single time point) without averaging, the separation is 740~$\mu s$. The time stamp indicate the moment after the plasma startup. The emissivity is shown in a colour pattern, which is scaled to the maximum in every image. This would simplify tracing profile evolution. The absolute intensity ramps up during first six frames and then ramps down as it seen in Figure~\ref{I_vs_t_46617}, where the total number of emitted D-$\alpha$ line  photons per second at corresponding time moments is shown. Thus, the brightest frame No.~6 has $\approx 100$ times more light than the first frame. In this confinement regime, deuterium plasma is created by the arc-discharge source located in the opposite (right) expander, see Figure~\ref{fig:gdt}. Eight 25~keV deuterium beams fire at 3.5~ms with the injection pulse duration of 5~ms. The D-$\alpha$ emissivity distribution remains sharp and relatively symmetric during first 4.4~ms after startup. Afterwards, the profile broads with a more evident asymmetry and an increasing shift to the left. Basically, this effect is consistent with the expander gas puff {\bf 12} in Figure~\ref{fig:w-exp_WIDA}. After the heating beam pulse is finished (frame No.~8 at $t=8.88$~ms and further), the emissivity profile evolves to a more compact shape with a gradually decreasing X-offset.

\begin{figure}[htbp]
\centering 
\includegraphics[width=.49\textwidth, origin=c]{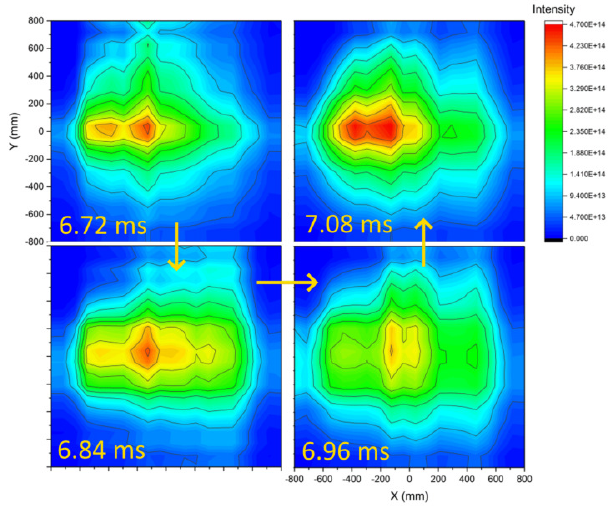}
\caption{\label{I_profiles_46612} Fast dynamics of  D-$\alpha$ emissivity profile during MHD event in GDT shot 46612.}
\end{figure}

Another set of reconstructed emissivity images in Figure~\ref{I_profiles_46612} taken in the GDT shot 46612, exhibits an example of fast dynamics associated with the MHD plasma instability during the neutral beams injection. Images are plotted with the time step of 120~$\mu s$ at the same intensity scale, which is shown in Figure~\ref{I_profiles_46612}. 

\section{Analysis of fluctuations}
\label{Correlations}
Evolution of plasma flow in the boundary region of the absorber indicate the presence of fluctuations. They appear as on oscillating signal component in Langmuir probes data, magnetic coils and measurements of line emission intensity as well. Typical edge LOS signals (see Figures~\ref{LOS_signals} and \ref{LOS_signals_scaled}) give evidence of the plasma turbulence that is probably expressed in fluctuations of electron and ion density. Some of transient events observed by the optical diagnostic, are clearly connected to low modes of MHD instability with azimuthal wavenumbers $m=0\div2$ (see for example Figure~\ref{I_profiles_46612} in the Section~\ref{I_profiles}). In many discharges, edge fluctuations of D-$\alpha$ emissivity reaches $\sim30 \%$. Basic analysis tools leaning on time domain correlations between signals and spatial coherence, can give a good insight into the characteristic frequency and wavenumber spectrum \cite{corran-1, corran-2, corran-3, corran-4}. Here we present a brief overview of the use case of D-$\alpha$ line intensity correlation measurements in the GDT expander plasma not delving into the underlying physics.

A similarity of two signals in time domain is defined by the cross-correlation function
\begin{equation*} 
\rho_{xy}(t) = \int_{-\infty}^{\infty}s_x^*(\tau)s_y(t+\tau)d\tau, 
\end{equation*}
where $s_x(t), s_y(t)$ are time traces (complex conjugate makes no difference in our case of real-valued data).
\begin{figure}[htbp]
\centering 
\includegraphics[width=.5\textwidth, origin=c]{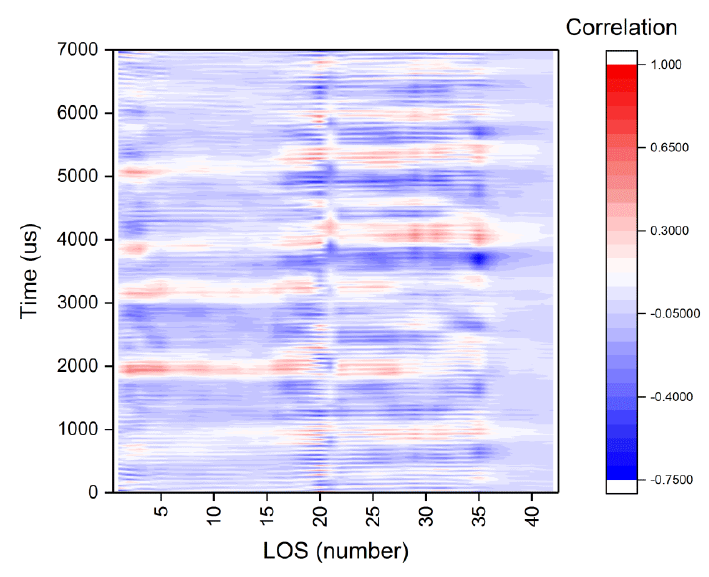}
\caption{\label{Corr_0_all_46933} Correlation between LOS-1 and other LOS in GDT shot 46933.}
\end{figure} 
Figure~\ref{Corr_0_all_46933} illustrates the example of 2-d correlation function $\rho_{(LOS-1)}(t, y)$ between the LOS-1 and the full set of LOS, $y=1\div42$. Two time scales are apparent in this periodic correlation function. The bigger frequency of $f_{MHD}\cong 14$~kHz is associated with the MHD interchange mode $m=1$, the smaller one of $f_{rot}\approx 1$~kHz corresponds to solid-body rotation of non-symmetric plasma column. These relatively small oscillation frequencies are typical for GDT regimes with the vortex confinement, where radial electric field in the plasma is controlled by biasing electrodes.

In order to further examine the turbulence in the frequency domain, the cross-power spectrum is computed according to the formula $P_{xy}(f) = \mathcal{F}(s_y)\mathcal{F^*}(s_x)$, where $ \mathcal{F}(s)$ denotes the signal Fourier transform. The function of coherence 
\begin{equation*}
\gamma_{xy}(f) = \frac{|P_{xy}(f)|^2}{P_{xx}(f)P_{yy}(f)}
\end{equation*}
estimates a stability of the cross-phase between two signals in time and space for the given frequency.
\begin{figure}[htbp]
	\centering
	\begin{subfigure}{.45\textwidth}
		\includegraphics[width=\textwidth]{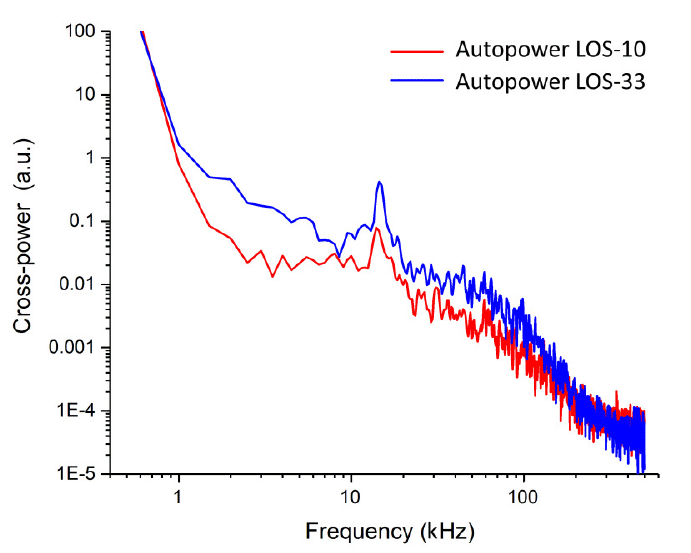}
		\caption{\label{Autopower} Autopower spectrum of signals LOS-10 (chord radius 64 mm) and LOS-33 (chord radius 495 mm).  }
	\end{subfigure}
	\qquad
	\begin{subfigure}{.45\textwidth}
		\includegraphics[width=\textwidth]{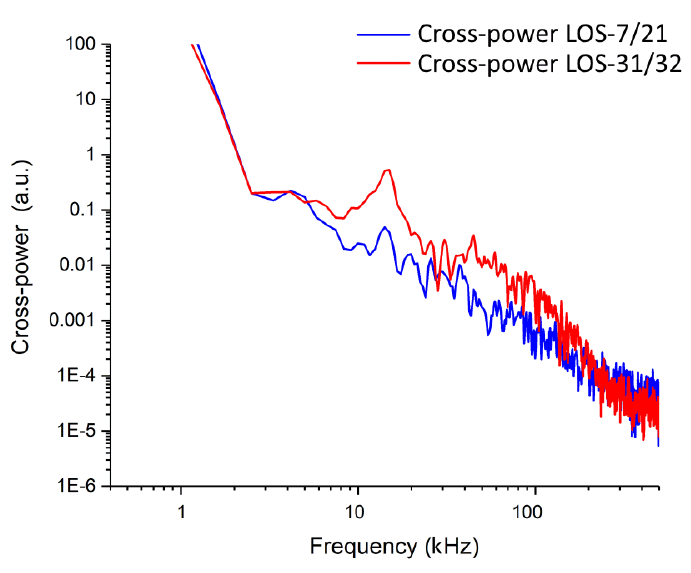}
		\caption{\label{Crosspower} Cross-power spectrum of signals LOS-7 and LOS-21 (chord radii 256 and 592 mm), LOS-31 and LOS-32 (chord radii 430 and 465 mm).  }
	\end{subfigure}
	\caption{Power spectrum of turbulence in shot 46933.}
\end{figure} 
Figure~\ref{Autopower} shows the autopower spectrum calculated by the signals in LOS-10 and LOS-33. Comparing the two plots, one may point out a remarkably larger fluctuation level on the plasma periphery than in the core (LOS-33 has the radius of 495~mm which projects to the limiter radius of 150~mm in the central plane coordinates -- the plasma boundary). The whole fluctuation spectrum spans over the range $1\div100$~kHz or even further with the prominent peak at the frequency of $\cong 14$~kHz, which is linked to large-scale MHD instability modes.
Same comparison between two curves in Figure~\ref{Crosspower} qualitatively demonstrates an observable degree of correlation in two neighbouring viewing chords at the plasma edge: LOS-31 and 32 with the radii of 430~mm and 465~mm over the spectral range of approximately  $5\div100$~kHz (red curve). A relative drop in the blue curve in Figure~\ref{Crosspower} witnesses that such correlations vanish between intermediate radii (LOS-7, 256~mm) and the edge (LOS-21, 592~mm).
\begin{figure}[htbp]
	\centering
	\begin{subfigure}{.45\textwidth}
		\includegraphics[width=\textwidth]{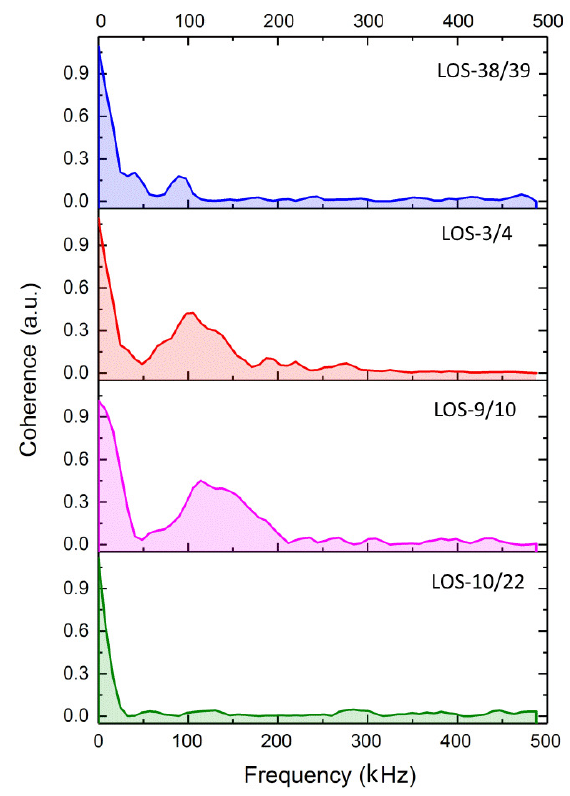}
		\caption{\label{Coherence} Coherence between four pairs of LOS.  }
	\end{subfigure}
	\qquad
	\begin{subfigure}{.46\textwidth}
		\includegraphics[width=\textwidth]{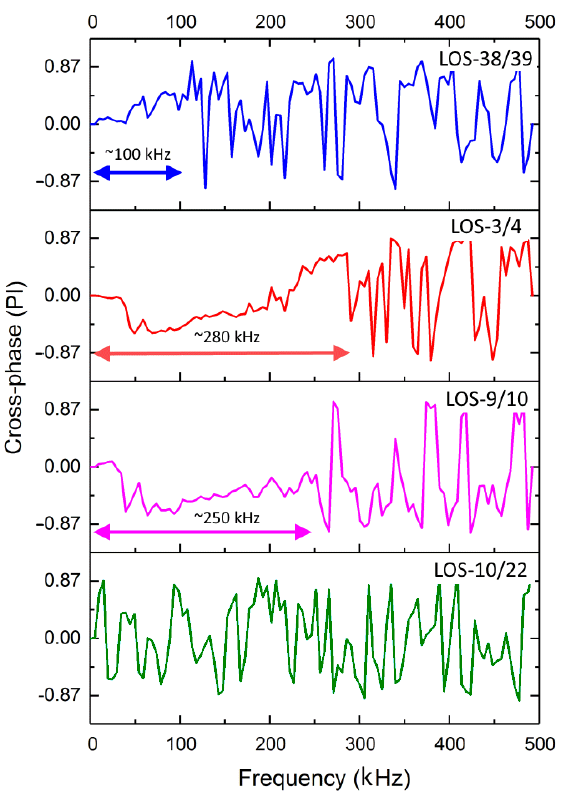}
		\caption{\label{Crossphase} Cross-phase between four pairs of LOS. }
	\end{subfigure}
	\caption{Illustration of spatial coherence of line-integrated intensity in different observation chords. }
\end{figure} 
To highlight the broadband frequency spectrum of the plasma turbulence, Figures~\ref{Coherence} and \ref{Crossphase} are provided with the linear axis scale. Coherency and cross-phase of four LOS pairs are shown: LOS-38 and LOS-39 ($r=630$~mm, $r=659$~mm -- edge); LOS-3 and LOS-4 ($r=481.5$~mm, $r=432.6$~mm -- mid-radii); LOS-9 and LOS-10 ($r=126$~mm, $r=64$~mm -- core); LOS-10 and LOS-22 ($r=64$~mm, $r=147.5$~mm -- two core LOS at  transverse directions). The latter function should be considered as the noise trace at frequencies above 50~kHz. The corresponding cross-phase function of LOS-10 and LOS-22 in Figure~\ref{Crossphase} behaves stohastically proving the absence of spatial correlations in the line-integrated emission acquired from those chords. On the contrary, other three combinations of LOS feature a large level of spatial coherence. The cross-phase for these spatial separations varies smoothly within the marked frequency range.

\section{Conclusion}
\label{conclusion}
The described optical tomography diagnostic is remaining at the commissioning stage at GDT. A robust modular design of the optical registration system and signal acquisition electronics allows for a gradual build-up of channels, while the system is already capable of delivering the valuable physical data. The nearest future objective is to introduce three more LOS beams similar to the installed bundle-1 (see Figure~\ref{fig:w-exp_WIDA}). The diagnostic is routinely working in experiments for the study of axial particle and energy transport in the axially symmetric gas dynamic trap. A part of this work package, is examination of a massive gas injection in the expander and its impact on the central cell electron temperature and plasma MHD stability. The preliminary results reported in this paper, encourage to expand the task list with the study of the discovered broadband plasma turbulence.

\acknowledgments
This work is supported by the Russian Science Foundation, project No.~18-72-10084 issued on 31.07.2018.

\end{document}